\documentstyle[pre,aps,preprint,12pt,times,graphicx,amssymb]{revtex}

\begin{document}
\draft   
\title{Restricted feedback control of one-dimensional maps}
\author{Kevin Hall$^{1,}$\footnote{email: hall@entelos.com} and David J. Christini$^{2,}$\footnote{email: dchristi@med.cornell.edu}}

\address{
$^{1}$
Entelos, Inc.,
Menlo Park, CA 94025
}
\address{
$^{2}$
Division of Cardiology,
Department of Medicine,\\
Weill Medical College of Cornell University,
New York, NY 10021
}

\date{\today}
\maketitle

\begin{abstract}

Dynamical control of biological systems is often restricted by the
practical constraint of unidirectional parameter perturbations.  We
show that such a restriction introduces surprising complexity to the
stability of one-dimensional map systems and can actually improve
controllability.  We present experimental cardiac control results that
support these analyses. Finally, we develop new control algorithms
that exploit the structure of the restricted-control stability zones
to automatically adapt the control feedback parameter and thereby
achieve improved robustness to noise and drifting system parameters.

\end{abstract}


\section{Introduction}

Recent success controlling complex dynamics of nonlinear physical and
chemical
systems~\cite{ditto:1990a,hunt:1991a,peng:1991a,petrov:1992a,roy:1992a,carroll:1992a,gills:1992a,schwartz:1993b,bielawski:1993a,gauthier:1994a,socolar:1994a,petrov:1994a,rulkov:1994a,colet:1994a,hubinger:1994a,christini:1996b}
has opened the door for the control of biological rhythms.  Some
researchers have speculated about the medical implications of
controlling heart-rate dynamics or brain
rhythms~\cite{garfinkel:1992a,schiff:1994b,weiss:1994a,glass:1996a,hall:1997a,christini:1999c}. However,
biological systems typically have characteristics that require special
consideration. For example, biological control studies to
date~\cite{garfinkel:1992a,glass:1994a,schiff:1994b,christini:1996a,watanabe:1996a,brandt:1996a,hall:1997a,christini:1997b,brandt:1997b,sauer:1997a,christini:2000a}
have required that the control interventions be unidirectional ---
only allowing shortening of a parameter.  Such a restriction is
somewhat analogous to trying to balance a broomstick vertically on
one's palm using horizontal hand movements in only one
direction. Intuitively, one might expect that such a restriction would
limit controllability. However, as we will demonstrate in this paper,
such a restriction introduces some surprising complexity to the
stability properties of controlled one-dimensional map systems.

In fact, the unidirectional restriction can actually improve the
controllability of some systems~\cite{fn:gauthier_ind}.  In this
paper, we will show how restricted control can introduce stability
zones that do not exist in the unrestricted case. Furthermore, we will
show that some of these zones were present in recent cardiac control
experiments~\cite{hall:1997a}.  Finally, we will exploit the structure
of the stability zones in robust new control algorithms that
automatically adapt the control feedback parameter.

\section{Delayed feedback control of systems described by one-dimensional maps}

In this study we will consider the control of systems whose dynamics can be
described by one-dimensional maps:
\begin{equation}
X_{n+1} = f(X_n, \lambda),
\label{eqn:map4}
\end{equation}
where $X_n$ is the variable to be controlled and $\lambda$ is an
experimentally accessible system parameter. The goal is to stabilize
the system state point $\xi_n=[X_{n},X_{n-1}]$ about an unstable
period-1 fixed point $\xi^*=[X^*,X^*]$, where $X^* = f(X^*, \lambda)$,
by perturbing $\lambda$ by an amount:
\begin{equation}
\delta \lambda_n = \frac{\alpha}{2} (X_{n-1} - X_n),
\label{eqn:deltalambda}
\end{equation}
where $\alpha$ is the feedback gain parameter.

The advantage of such a control scheme is that relatively little
\emph{a priori} system information is required to
implement control and stabilize $\xi^*$. The only requirement is
knowledge of the sign of $\frac{\partial f}{\partial
\lambda}$ so that perturbations can be applied in the correct
direction.  In fact, knowledge of the value of $\xi^*$ is unnecessary
because the controlled system's fixed point is identical to that of
the uncontrolled system. Furthermore, if the fixed point drifts during
the course of the control (as is common for biological systems), the
controlled system will track the fixed point, provided that the system
stays in the stable range of the feedback gain parameter $\alpha$.

The control algorithm of Eq.~\ref{eqn:deltalambda} is an example of
delayed feedback control, a technique that has been used in a variety
of modeling and experimental
studies~\cite{hunt:1991a,peng:1991a,petrov:1992a,pyragas:1992a,gauthier:1994a,christini:1996a}.
In section~\ref{sec:AVNexperiments} we will present an example of a
biological system with constraints that restrict the control algorithm
--- only allowing unidirectional perturbations of $\lambda$. The
purpose of this study is to examine the implications of such a
restriction.

\subsection{Linear stability analysis of unrestricted delayed feedback control}

For unrestricted control, in which $\delta \lambda_n$ can be positive
or negative, linearizing the controlled system about a fixed point at
the origin gives:
\begin{eqnarray}
X_{n+1} & = & A X_n + \beta (Y_n - X_n), \label{eqn:lin}\\
Y_{n+1} & = & X_n, \nonumber
\end{eqnarray}
where $A \equiv \frac{\partial f}{\partial X}|_{\xi^*}$ and $\beta \equiv
\frac{\alpha}{2}\bigl( \frac{\partial f}{\partial \lambda}
\bigr)|_{\xi^*}$.

The eigenvalues of Eq.~\ref{eqn:lin} are $\bigl(A - \beta \pm \sqrt{(A -
\beta)^2  + 4  \beta} \bigr)/2$.   The fixed point  is stable provided
that both eigenvalues fall inside  the unit circle.  This condition is
met when:
\begin{equation}
-1 < \beta < \frac{1}{2}(A + 1),
\label{eqn:unrestzone}
\end{equation}
for $A < 1$~\cite{hall:1997a,ushio:1996a}.  Note that the stability
zone shrinks to zero for $A \leq -3$, thereby limiting the
applicability of the unrestricted control algorithm to maps with a
sufficiently shallow slope (i.e., $-3 \leq A <1$) at
$\xi^*$. Furthermore, for $A > 1$ there exists no real value for
$\beta$ such that the eigenvalues fall within the unit circle. Thus,
unstable positively-sloped fixed points cannot be stabilized.

\section{Restricted delayed feedback control}

Restricting the above control algorithm by only allowing shortening of
$\lambda$ gives the following controller:
\begin{eqnarray}
\delta \lambda_n = \Theta_n \frac{\alpha}{2} (X_{n-1} - X_n),
\label{eqn:restrictdeltalambda}
\end{eqnarray}
where

\begin{eqnarray}
\Theta_n & = &
\left\{
    \begin{array}{ll}
       1 & \;\; \hbox{if } (X_n - Y_n) > 0,   \\
       0 & \;\; \hbox{otherwise}.
    \end{array}
\right.
\label{eqn:cond}
\end{eqnarray}
\noindent
Thus, when $\Theta_n = 1$ the control is active (i.e., a perturbation
is delivered), and when $\Theta_n = 0$ the control is inactive (i.e.,
no perturbation is given).

\subsection{Linear stability analysis of restricted delayed feedback control}

The restricted control algorithm of Eq.~\ref{eqn:restrictdeltalambda}
gives the following linearized controlled system:
\begin{eqnarray}
X_{n+1} &  = & A X_n +  \Theta_n \beta (Y_n  - X_n), \\
\label{eqn:restlin}
Y_{n+1} & = & X_n. \nonumber
\end{eqnarray}
Geometrically, the restriction of Eq.~\ref{eqn:cond} means that
perturbations will only be applied if the state point $\xi_n$ lies
above the return-map line of identity $X_{n+1} = X_n$. The dynamical
effects of this restriction depend on the sign of the slope at
$\xi^*$.

\subsubsection{Restricted control for $A < -1$}
\label{sec:control_a_below_minus1}

Typical dynamics of restricted control for negatively-sloped unstable
directions ($A<-1$) are depicted in
Fig~\ref{fig:x_return_a_below_minus1}. This figure shows eight control
trials of a linear map with $A = -4$ for different values of $\beta$;
there are four examples of stable control and four examples of
unstable control.

Figure~\ref{fig:x_return_a_below_minus1}(a) shows a case in which the
restricted control algorithm failed to stabilize the unstable fixed
point $\xi^*$ [which is located at the intersection of the
uncontrolled system map (solid line) and the line of identity (dotted
line), and is denoted by a solid triangle] with $\beta=-2.80$. The
dot-dash lines correspond to the system maps when $\lambda$ is
perturbed. A series of arrows originate at the initial state point and
connect consecutive state points (solid circles, numbered
consecutively). In this case, the initial state point ($1$) is
followed by a control intervention, which causes the next iterate
($2$) to fall below the line of identity. According to
Eq.~\ref{eqn:cond}, the next iterate ($3$) will be uncontrolled and
therefore will fall on the solid line. Furthermore, because the first
controlled iterate ($2$) was less than the fixed point, the next
iterate ($3$) will be above the line of identity, leading to a control
intervention at the fourth iterate ($4$).  Thus, control is applied every
other iterate so that the sequence of $\Theta_n$ is 0101.... In this
case, the fixed point is not stabilized because control fails to
direct the system closer to the fixed point (i.e., the arrows spiral
away from $\xi^*$).

Figure~\ref{fig:x_return_a_below_minus1}(b) shows control with the
same value of $A$, but using a slightly more negative $\beta$ value,
$\beta=-3.1$. For these parameter settings, it can be seen that
control is also applied every other iterate so that the sequence of
$\Theta_n$ is again 0101.... However in this case, the fixed point is
stabilized successfully because the control interventions direct the
system closer to the fixed point (i.e., the arrows spiral towards
$\xi^*$). Note that in order to maximize the clarity of the control
sequence diagram, the axes in panel (a) are not scaled the same as
those for panel (b). Similarly, for all of
Fig.~\ref{fig:x_return_a_below_minus1}, axes from different panels are
not necessarily scaled the same.

Figure~\ref{fig:x_return_a_below_minus1}(c) shows a trial in which
$\beta$ was decreased to $-3.23$.  As in the previous example, the
first controlled iterate ($3$) is below the line of identity
(dictating that the next iterate ($4$) is uncontrolled). However, in
this case the perturbation is larger than would occur with the
parameter settings of Fig.~\ref{fig:x_return_a_below_minus1}(b), such
that the controlled iterate ($3$) is slightly larger than the fixed
point.  This dictates that the next iterate ($4$) is below the line of
identity, which leads to a second consecutive uncontrolled iterate
($5$). Thus, control is applied in a 001001... sequence.  This
sequence is stable for $\beta=-3.23$ because the control perturbations
direct the system closer to the fixed point. However, if $\beta$ is
decreased to $-3.40$, it can be seen from
Fig.~\ref{fig:x_return_a_below_minus1}(d) that this generates a
001001... control sequence that is unstable because the system is directed
away from the fixed point.

If $\beta$ is made more
negative, then a new control sequence is achieved.
Figure~\ref{fig:x_return_a_below_minus1}(e) shows an unstable
011011... sequence for $\beta=-5.50$.  In this case, the first
perturbation is so large that the controlled iterate ($2$) is above
the line of identity, dictating that the next iterate ($3$) is also
controlled.  The second controlled iterate ($3$) is below the line of
identity and below the fixed point, thereby producing the
011011... sequence.  In this case, the fixed point is not stabilized
because the control perturbations moved the state point away from the
fixed point. However, when $\beta$ is decreased to $-5.76$, as seen in
Fig.~\ref{fig:x_return_a_below_minus1}(f), a 011011... sequence
stabilizes the fixed point.

Like the transition from the stable 0101... sequence to the stable
001001... sequence depicted in
Figs.~\ref{fig:x_return_a_below_minus1}(b) and (c), there is a
transition from the stable 011011... sequence to the stable
00110011... sequence as $\beta$ is decreased further. Figure
\ref{fig:x_return_a_below_minus1}(g)  shows the stable 00110011... case  for
$\beta=-5.798$. The 00110011... sequence becomes unstable as $\beta$
is decreased still further;
Figure~\ref{fig:x_return_a_below_minus1}(h) depicts this case for
$\beta=-5.95$.

The progression of unstable and stable periodic control sequences
continues indefinitely as $\beta$ is decreased.  In fact, the
switching parameter $\Theta_n$ imposes the following progression of
control sequences as $\beta$ is decreased from zero: unstable $01^1$,
stable $01^1$, stable $001^1$, unstable $001^1$, unstable $01^2$,
stable $01^2$, stable $001^2$, unstable $001^2$, ..., unstable
$01^\infty$, stable $01^\infty$, stable $001^\infty$, unstable
$001^\infty$, where $1^k$ denotes $k$ consecutive ones (control
perturbations) before the sequence repeats itself.

Because of the progression of the control sequences imposed by the
switching term $\Theta_n$, $X_{k+1}$ can be expressed as:
\begin{equation}
X_{k+1} = e_k X_0,
\label{eqn:x_k+1}
\end{equation}
where $e_k$ is given by the following iterative expression:
\begin{equation}
e_k = (A - \beta) e_{k-1} + \beta e_{k-2},
\label{eqn:e_k}
\end{equation}
with $e_0 = A$ for all sequences and $e_1 = A^2 + \beta (1 - A)$ for
the $01^k$ sequences or $e_1 = A^2$ for the $001^k$ sequences.

The boundaries of the stability zones can be computed by using the
criterion that stable sequences move the system closer to the fixed
point after one control sequence. Because $X_{k+1}$ is the last
iterate of the first $01^k$ control sequence and $X_{k+2}$ is the last
iterate of the first $001^k$ sequence, the stability conditions are
$e_k < 1$ and $e_{k+1} < 1$ for the $01^k$ and $001^k$ sequences,
respectively.  Therefore, the boundaries are given by $k$ degree
polynomials in $\beta$. For example, the $k=1$ control sequences are
stable for $1 + A + 1/A \leq \beta \leq 1+A$ for $A <
-1$~\cite{bielawski:1993a}.  Figure
\ref{fig:beta_vs_a_below_minus1} depicts the stability zones (shaded
regions) for $k = 1$ and $k = 2$.  The boundaries between the stable
$01^k$ and $001^k$ sequences are defined by the condition $e_k = 0$
(dotted curves in Fig.~\ref{fig:beta_vs_a_below_minus1}).  These
curves mark the optimal parameter values for a given stability region,
because the fixed point is reached after a single control sequence
$01^k$.

The striking feature of this analysis is that the domain of control is
extended by the restriction of
Eq.~\ref{eqn:cond}~\cite{fn:gauthier_ind}.  In fact,
Fig.~\ref{fig:beta_vs_a_below_minus1} shows that for all $A<-1$,
stable control sequences exist for the restricted system. This is in
contrast with the unrestricted system, for which stable control
sequences exist only for $A>-3$, as shown by the dashed triangular
region marking its stability zone (according to
Eq.~\ref{eqn:unrestzone}).

While there are an infinite number of stable zones corresponding to an
arbitrary number $k$ consecutive control perturbations, the stability
zones are bounded by the curve $\beta = A - 2 - 2 \sqrt{1 - A}$. This
boundary is computed by recognizing that as $k$ approaches infinity,
control is always active and $X_{n+1} > X_n$ for every iterate.  Thus,
the algorithm behaves just like the unrestricted control of
Eq.~\ref{eqn:lin} with real eigenvalues greater than one --- a
condition met only when $\beta$ is below the boundary.

\subsubsection{Restricted control for $A > 1$}
\label{sec:control_a_above_1}

Typical dynamics of restricted control for positively-sloped unstable
directions ($A>1$) are depicted in
Fig~\ref{fig:x_return_a_above_1}. This figure shows four control
trials of a linear map, with $A = 2.1$, for different values of $\beta$;
there are three examples of unstable control and one example where
$\xi^*$ is controlled.

Figure~\ref{fig:x_return_a_above_1}(a) shows an unstable
$01^\infty$ sequence for $\beta = 1.5$.  In this case, the perturbations
are so small that all state points lie above the line of identity. The
control slows the exponential growth (which would be marked by a rapid
exit from $\xi^*$ along the solid line), but fails to force an
approach to $\xi^*$.

If $\beta$ is decreased to $\beta = 2.5$, then the
perturbations are large enough so that the first controlled $X_n$ is
smaller than the previous $X_{n-1}$. $\xi_n$ then falls below the
line of identity, thereby generating the $01^1$ control sequence
depicted in Fig.~\ref{fig:x_return_a_above_1}(b). In this case, the
sequence is unstable because a given controlled point is further from
$\xi^*$ than the previous controlled point.

For $\beta = 3.5$, control is successful; the state point approaches
$\xi^*$ in a $01^1$ sequence as shown in
Fig.~\ref{fig:x_return_a_above_1}(c). Such control is successful for a
noise-free model system. However, for a real-world system, once the
state point is sufficiently close to $\xi^*$, noise will eventually
kick $\xi_n$ to the opposite side of $\xi^*$. Subsequent $\xi_n$ will
fall below the line of identity, causing the control to be deactivated
and leading to an exponential departure from $\xi^*$.

When $\beta$ is increased further, the first perturbation can be so
large that the state point will be kicked to the other side of $\xi^*$
as shown in Fig.~\ref{fig:x_return_a_above_1}(d) for $\beta = 4.5$.
Again, control is subsequently deactivated and the system diverges
from the fixed point.

The boundaries between the different control sequences are depicted in
Fig.~\ref{fig:beta_vs_a_above_1}. The unstable $01^\infty$ sequence
occurs for $\beta<A$. This boundary was computed by finding the value
for $\beta$ such that the first controlled iterate falls on the line
of identity; this occurs for $\beta = A$. For $\beta > A$, the first
controlled iterate lies below the line of identity, thereby shutting
the control off for the next iterate. Thus, $\beta = A$ marks the
boundary between the $01^\infty$ and $01^1$ sequences.  The boundary
between unstable and stable $01^1$ sequences was computed by finding
the value for $\beta$ such that the iterate subsequent to control is
equal to the previous uncontrolled iterate; this occurs for $\beta =
1+A$.  For slightly larger $\beta$ values, the iterate subsequent to
control is closer to $\xi^*$ than the previous uncontrolled
iterate. Thus, the unstable $01^1$ sequence is bounded by $A < \beta <
1 + A$. The final boundary marks the end of the converging $01^1$
sequence and can be found by determining the value for $\beta$ such
that the first controlled iterate lands at $\xi^*$. This occurs for
$\beta = A^2/(A-1)$. For $\beta > A^2/(A-1)$, the first controlled
iterate, and all subsequent iterates, lie below the line of identity
thereby shutting of the control. Therefore, the converging $01^1$
sequence occurs in the shaded region $1 + A < \beta <
A^2/(A-1)$~\cite{bielawski:1993a,fn:bielawski}, and the unstable $010^\infty$
sequence occurs for $\beta > A^2/(A-1)$.

Thus, for $A > 1$ the best that the restricted control algorithm can
offer is a temporary reversal of divergence from the fixed
point. However, in section~\ref{sec:auto_adapt} we will make a simple
modification to the restricted control algorithm for $A>1$ that will
keep the system in the vicinity of $\xi^*$ in the presence of noise.

\section{Experimental observation of restricted control sequences}
\label{sec:AVNexperiments}

As described in Ref.~\cite{hall:1997a}, we have studied the control of a
particular cardiac conduction interval, known as the atrioventricular
(AV) nodal conduction time, in \emph{in vitro} rabbit heart
experiments. Because of the nonlinear excitation properties
of AV-nodal tissue, the dynamics of AV-nodal conduction can bifurcate
from a period-1 regime (where every impulse propagates through the
AV-node at the same rate) to a period-2 regime (where propagation time
alternates in a long, short, long, etc. pattern on a beat-to-beat
basis) during rapid atrial excitation. It has been
demonstrated~\cite{sun:1995a,amellal:1996a} that these dynamics can be
described by a period-doubling bifurcation of a one-dimensional map of
the form of Eq.~\ref{eqn:map4} where
$X_n$ is the AV-nodal conduction time and $\lambda_n$ is the time
between when the AV-node finishes conducting one impulse and when it
starts conducting the next.

The goal in Ref.~\cite{hall:1997a} was to eliminate the alternating
rhythm by stabilizing the underlying period one fixed point
$X^*$. This was achieved by delivering electrical stimuli to the
atrial tissue in order to transiently shorten $\lambda_n$. Because
there is no practical way to lengthen $\lambda_n$, the timing of the
electrical stimuli was determined by the restricted controller
of Eq.~\ref{eqn:restrictdeltalambda}.

Although the \emph{in vitro} rabbit cardiac system of
Ref.~\cite{hall:1997a} was not linear, application of the restricted
control algorithm resulted in several of the control sequences
predicted in the above linear system for $A < -1$. These control
sequences were especially clear at the initiation of control when
perturbations were largest.

For example, Fig.~\ref{fig:hall_97a_fig}(a) shows the variable $X_n$
and the control parameter $\lambda_n$ during an unstable $01^1$ sequence
for a feedback gain $\alpha = 3.3$ (corresponding to a negative
$\beta$ because $\frac{\partial f}{\partial \lambda} < 0$
in the cardiac control experiments).  The first controlled beat is
indicated by the arrow and corresponds to a negative perturbation of
$\lambda_0$ (all control perturbations are negative as imposed by the
switching term $\Theta_n$). Because the system was nonlinear,
oscillatory growth of $X_n$ was quenched and the original large
amplitude alternation of $X_n$ was reduced in magnitude --- but not
eliminated.

When $\alpha$ was increased to $5.0$ (as shown in
Fig.~\ref{fig:hall_97a_fig}(b); corresponding to a later segment of
the same control trial that is shown in
Fig.~\ref{fig:hall_97a_fig}(a)), the system shifted to a stable
$001^1$ sequence that eliminated the alternation of $X_n$. After the
fourth perturbation to $\lambda_n$ (beat 303), the system shifted to a
stable $01^1$ sequence. This shift resulted from the close proximity
of the $01^1$ and $001^1$ stable zones
(Fig.~\ref{fig:beta_vs_a_below_minus1}); noise or drift in the system
can cause such transitions.

Figure~\ref{fig:hall_97a_fig}(c) shows a stable $001^2$ control
sequence that eliminated the alternation of $X_n$ in a different
rabbit heart using $\alpha = 2.5$.  (Note that the $\alpha$ values
from distinct trials are independent.) Similar to the sequence
transitions in Fig.~\ref{fig:hall_97a_fig}(b), the system switched to
its adjacent stable $01^2$ control sequence shortly after the control
was initiated, and later switched back to the stable $001^2$ control
sequence.

\section{Modifications to the restricted control algorithm}

\subsection{Automatic adaptation of the feedback gain for $A < -1$}
\label{sec:auto_adapt}

Figure~\ref{fig:hall_97a_fig}(a) (unsuccessful) and (b) (successful)
demonstrate that successful control is dependent on the proper choice
of $\alpha$. Such dependence is a critical limitation given that the
information required to determine the correct value of $\alpha$ ($A$,
$\xi^*$, and $\frac{\partial f}{\partial X}|_{\xi^*}$) cannot be
easily determined prior to control. Furthermore, the nonstationarities
typical of biological systems imply that the appropriate value of
$\alpha$ may drift over time, thereby increasing the likelihood of
control failure if $\alpha$ is fixed. To eliminate the limitations of
a fixed $\alpha$ value chosen prior to control, we have developed a
new technique that adaptively estimates
$\alpha$~\cite{fn:previous_adaptive}.  This adaptive approach is
especially appropriate for applications (e.g., cardiac arrhythmia
control) that cannot afford control failure of the type shown in the
control attempt of Fig.~\ref{fig:hall_97a_fig}(a).

This new technique exploits the structure of the stability zones to
automatically adapt $\alpha$ such that $\xi^*$ is stabilized more
robustly.  Because multiple perturbations away from the fixed point
are not desirable, the optimal stability zone is the $k = 1$
zone. Furthermore, because the stable $k = 1$ zone has the largest
area, it will be the most robust to noise and drifting system
parameters. To target this zone, $\alpha$ is adapted according to:

\begin{eqnarray}
\alpha_n & = &
\left\{
    \begin{array}{ll}
       \alpha_{n-1}  + \delta \alpha & \;\; \hbox{if $\Theta_{n-4}$ ... $\Theta_{n-1}$ = 0101  or 1010},\\
       \alpha_{n-1}  - \delta \alpha & \;\; \hbox{otherwise},
    \end{array}
\right.
\label{eqn:adapt_alpha}
\end{eqnarray}
where $\delta \alpha$ is a small increment.  When
$\bigl(\frac{\partial f}{\partial \lambda}\bigr)|_{\xi^*}$ is negative
(as in the cardiac experiments of Ref.~\cite{hall:1997a}), $\alpha$ and
$\delta \alpha$ are positive. Otherwise they are negative.

The algorithm of Eq.~\ref{eqn:adapt_alpha} is motivated by examining the
stability zones in Fig.~\ref{fig:beta_vs_a_below_minus1}.  For $k=1$,
optimal control occurs when $\beta$ is at the boundary between stable
$01^1$ and stable $001^1$ (dotted curve of
Fig.~\ref{fig:beta_vs_a_below_minus1}). If $\beta$ is too large, the
control sequence will be $01^1$ (unstable if $\beta$ is so large that it
is above the k=1 stability region or stable if $\beta$ is within the
stability region but above the optimal control boundary). In such a
case, the adaptation of Eq.~\ref{eqn:adapt_alpha} will decrease
$\beta$. In contrast, if $\beta$ is too small, the control sequence
will be $001^1$ or some sequence with $k > 1$ (unstable if $\beta$
is so small that it is below the $k=1$
stability region or stable if $\beta$ is within the stability region
but below the optimal control boundary).  In such a case, the
adaptation of Eq.~\ref{eqn:adapt_alpha} will increase $\beta$. Thus, the
adaptation will adjust the system so that it oscillates between the
stable $01^1$ and stable $001^1$ sequences, provided that the increment
$\delta \alpha$ is small enough so that the stepsize of $\beta$ is
sufficiently less than the height of the $k=1$ stability
zone. Specifically, the condition $|\delta \alpha| < |A \bigl(\frac{\partial
f}{\partial \lambda}\bigr)|_{\xi^*}|^{-1}$ ensures that the stepsize is
less than half the height of the zone.

To illustrate the adaptive algorithm, we implemented the restricted
controller of Eq.~\ref{eqn:restrictdeltalambda}
with the feedback gain $\alpha$ replaced by $\alpha_n$ given by
Eq.~\ref{eqn:adapt_alpha}.  $\alpha_0$ was randomly chosen between $-5$
and $-10$ and $\delta \alpha = - 0.1$.  We applied this controller to
the quadratic map:
\begin{equation}
X_{n+1} = \lambda_n X_n (1 - X_n) + \zeta_n,
\label{eqn:quadmap}
\end{equation}
where $\zeta_n$ is a normally-distributed random variable with a mean
of zero and a variance of 0.001. The goal was to stabilize the fixed
point $X^* = (\lambda_0 -1)/\lambda_0$. Note that for the quadratic
map, $\bigl(\frac{\partial f}{\partial \lambda}\bigr)|_{\xi^*} =
\frac{\lambda - 1}{\lambda^2}$ is positive. Thus, the sign of the
perturbations is opposite to that of the cardiac experiments of
Ref.~\cite{hall:1997a}.

Figure~\ref{fig:quadcon} shows the results of an adaptive control
trial of the quadratic map.  For $1\leq n \leq 500$, $\lambda_0 =
3.30$, corresponding to an uncontrolled stable period-2 orbit.
Control was initiated at iterate $n=125$ with an arbitrary initial
value of $\alpha_n=-5.25$.  The adaptive algorithm rapidly stabilized
$\xi^*$ as the adaptations of $\alpha_n$ kept the system in the $k=1$
stability zone. The control was deactivated at $n = 375$ and the
period-2 cycle returned.

At $n = 500$, $\lambda_0$ was switched to $3.52$, which corresponds to a
period-4 rhythm.  Control was reactivated at $n = 625$ with an
arbitrary initial value of $\alpha_n = -8.85$.  The adaptive control
stabilized $\xi^*$ until the control was turned off at iterate 875.
At $n = 1000$, $\lambda_0$ was switched to 3.65, which is in the
chaotic regime.  Control was reactivated at $n = 1125$ with an
arbitrary initial value of $\alpha_n = -5.63$. $\xi^*$ was stabilized
after approximately 160 iterates. When control was deactivated at $n =
1375$ the chaos resumed. The oscillations in $\alpha_n$, as governed
by Eq.~\ref{eqn:adapt_alpha}, are apparent in each control phase of
the trial.

To demonstrate the ability of the adaptive algorithm to track a
drifting fixed point, we applied the control algorithm to a quadratic
map with $\lambda_0=3.0$ and $\lambda_n$ increased by an increment of
0.001 every iterate ($\lambda_{n+1}=\lambda_n+0.001$). As seen in
Fig.~\ref{fig:quadrift}, the small increments to $\lambda_0$
introduced a slow drift in the system dynamics and fixed
point. Control was initiated at iterate 250 while the system was in
its period-2 regime. Control was maintained for 500 iterates. During
this period, $X^*$ drifted from $X^*=0.692$ to $X^*=0.736$.
Nevertheless, the control algorithm had no trouble tracking $X^*$. In
fact, it can be seen that when control was deactivated at $n=750$, the
system had passed into the chaotic regime, an occurrence that did not
disrupt control. However, if the feedback gain was held fixed rather
than adapted, then control could not have been maintained for the
entire control period (not shown).

\subsection{Control of fixed points for $A > 1$}

In section~\ref{sec:control_a_above_1}, we demonstrated that the
restricted control algorithm can induce a transient approach towards
$\xi^*$ when $A > 1$. However, as mentioned, if the system is kicked
to the other side of $\xi^*$, control is deactivated and the system
diverges from $\xi^*$. If the algorithm could detect such an
occurrence and reverse the sign of the perturbations, then $\xi^*$
could be approached from the opposite side of
$\xi^*$~\cite{fn:unidirec_violation}.  This idea motivates the
following modification of the restricted control algorithm:
\begin{eqnarray}
\delta \lambda_n = \widehat{\Theta}_n \frac{\alpha_n}{2} (X_{n-1} - X_n),
\label{eqn:poscon}
\end{eqnarray}
where

\begin{eqnarray}
\widehat{\Theta}_n & = &
\left\{
    \begin{array}{ll}
       1 & \;\; \hbox{if $\Phi_n (X_n - Y_n) > 0$},\\
       0 & \;\; \hbox{otherwise},
    \end{array}
\right.
\label{eqn:Thetahat}
\end{eqnarray}
and
\begin{eqnarray}
\Phi_n & = &
\left\{
    \begin{array}{ll}
       -1 & \;\; \hbox{if $\Theta_{n-4}$ ... $\Theta_{n-1}$ = 1000},\\
       1 & \;\; \hbox{otherwise}.
    \end{array}
\right.
\label{eqn:Phi}
\end{eqnarray}

For stable control, a fixed value of the feedback gain ($\alpha_n =
\alpha_0$) is chosen so that $1 + A < \beta < A^2/(A-1)$.  In order to
adaptively control fixed points with $A > 1$, the controller in
Eq.~\ref{eqn:poscon} can be used with a modified adaptive feedback
gain algorithm:
\begin{eqnarray}
\alpha_n & = &
\left\{
    \begin{array}{ll}
       \alpha_{n-1}  + \delta \alpha & \;\; \hbox{if $\widehat{\Theta}_{n-4}$ ... $\widehat{\Theta}_{n-1}$ = 0101  or 1010},\\
       \alpha_{n-1}  - \delta \alpha & \;\; \hbox{otherwise}.
    \end{array}
\right.
\label{eqn:posadapt_alpha}
\end{eqnarray}
Such a combination is feasible because the control-sequence boundaries
for $A > 1$ dictate that Eq.~\ref{eqn:posadapt_alpha} will direct the
system towards the converging 0101... sequence (see
Fig.~\ref{fig:beta_vs_a_above_1}).  However, as in the case when
$A<-1$, the increment in the feedback gain should be chosen such that
$\alpha_n$ remains in the stable 0101... zone. Choosing $|\delta
\alpha| < |(A-1)\bigl(\frac{\partial f}{\partial
\lambda}\bigr)|_{\xi^*}|^{-1}$ ensures that the increment is less than half of
the height of the zone.

In order to illustrate the control of an unstable fixed point with
$A > 1$, we applied the modified control algorithm to the cubic map:
\begin{equation}
X_{n+1} = -4 (m + 1)X_n^3 + 6 (m + 1)X_n^2 - (2m + 3)X_n + \lambda_n + \zeta_n,
\label{eqn:cubemap}
\end{equation}
where $\lambda_n$ is perturbed according to Eq.~\ref{eqn:poscon}
with $\lambda_0 = 1$, $m$ is the slope of the map at the fixed point
$X^* = 0.5$, and $\zeta_n$ is a normally-distributed random variable with
a mean of zero and a variance of 0.001.

Figure~\ref{fig:cubecon} illustrates control of the cubic map without
adaptation of the feedback gain ($\alpha_n$ is fixed at $\alpha_0 = 8.0$).
For $1\leq n \leq 500$, $m = 2.2$, corresponding to an uncontrolled stable
period-2 orbit. After control was initiated at $n = 125$, $X^*$ was
stabilized after about 80 iterates. Stabilization of $X^*$ was
maintained until control was deactivated at $n = 375$, after which the
system returned to the period-2 orbit. At $n = 500$, $m$ was set to $2.7$,
which moved the system into the chaotic regime. Control was initiated
at $n = 625$. After about 10 iterates, $X^*$ was controlled for
about 20 iterates. The system subsequently escaped control for about
30 iterates before control was recaptured and maintained until the
algorithm was deactivated at $n = 875$.

The adaptive feedback gain algorithm is illustrated in
Fig.~\ref{fig:cubedrift}, which shows control of a drifting cubic map
with $m_{n+1} = m_n + 0.001$ and $m_0 = 2.0$.  Control was activated for
$250 < n < 750$. The fixed point was controlled successfully during
this period. The fixed point location does not change for the drifting
cubic map, but the system drifts into the chaotic regime by the end of
the control period.

\section{Conclusion}

Surprisingly, the typical biological restriction of unidirectional
control perturbations enhances the controllability of fixed points
with $A < -1$ in systems described by one-dimensional maps. Because
biological systems typically drift over time, dynamic control
algorithms must adapt to system nonstationarities. Although the
restricted delayed feedback control technique allows for moderate
tracking of the fixed point as long as the system remains within a
stability zone, it is ill-suited for systems with significant
drift. For such systems, automatic adaptation of the feedback gain
parameter ensures that the drifting system is directed to, and remains
within, the largest stability zone.  Thus, with the dual benefits of
the increased stability of unidirectional restricted control and the
adaptability of on-the-fly gain estimation, such control techniques
could be of significant value to the control of biological
systems. Indeed, a recent set of clinical
experiments~\cite{christini:2000c} have shown that adaptive restricted
control of this type can successfully eliminate the same alternating
AV-nodal conduction rhythm that was controlled in the rabbit
experiments of Ref.~\cite{hall:1997a}. Furthermore, we have shown that
simple modifications of the restricted control algorithm can control
fixed points with $A > 1$ --- an impossible task for the unrestricted
feedback controller. Thus, this algorithm may also have applicability
in physical systems that allow bidirectional perturbations.

\section*{ACKNOWLEDGMENTS}
\noindent
This work was supported, in part, by a grant from the American Heart
Association (0030028N) [DJC].

\newpage

\bibliographystyle{prsty} 	

\begin{thebibliography}{10}
\bibitem{fn:unidirec_violation}
Note that this modification of the algorithm violates the biologically
  motivated restriction of unidirectional perturbations. Nevertheless, not all
  systems are subject to such restrictions.
\end{thebibliography}


\newpage

\begin{figure}[!ht]

	\caption{Return maps showing the progression of control
	sequences as $A$ is decreased below $-1$. Sequential state
	points are numbered, the dotted diagonal line is the identity
	line $X_{n+1} = X_n$, the solid line is the map of the
	uncontrolled system with slope $A = -4$, the fixed point
	$\xi^*$ is denoted by the solid triangle, and the dot-dash
	lines show the system map when perturbed by control
	interventions. (a) $\beta=-2.8$ results in an unstable $01^1$
	control sequence. (b) $\beta=-3.1$; stable $01^1$. (c)
	$\beta=-3.23$; stable $001^1$. (d) $\beta=-3.4$; unstable
	$001^1$.  (e) $\beta=-5.5$; unstable $01^2$. (f)
	$\beta=-5.76$; stable $01^2$. (g) $\beta=-5.798$; stable
	$001^2$. (h) $\beta=-5.95$; unstable $001^2$. Note that axes
	from different panels are not necessarily scaled the same.}

\label{fig:x_return_a_below_minus1}
\end{figure}

\begin{figure}[!ht]

	\caption{Stability zones of unrestricted (Eq.~\ref{eqn:lin})
	and restricted (Eq.~\ref{eqn:restlin}) control for $A<-1$. The
	triangular region enclosed by the dashed lines in the upper
	right corner is the stability zone for unrestricted
	control. For restricted control, the $k=1$ ($01^1$ and
	$001^1$) and $k=2$ ($01^2$ and $001^2$) stability zones are
	the shaded regions enclosed by the solid curves (which are
	$k$-degree polynomials in $\beta$, determined via
	Eq.~\ref{eqn:e_k}, as described in the text). The dotted
	curves inside the zones mark the transition from $01^k$ to
	$001^k$. The annotated open circles \textbf{a}--\textbf{f}
	correspond to the control parameters for panels (a)--(f) of
	Fig.~\ref{fig:x_return_a_below_minus1}. The three
	vertically-spaced solid dots indicate that there are an
	infinite number of stability zones for larger $k$. The
	infinite sequence of stability zones is bounded by the curve
	$\beta = A-2-2\sqrt{1-A}$.}

\label{fig:beta_vs_a_below_minus1}
\end{figure}

\begin{figure}[!ht]

	\caption{Return maps showing the progression of control
	sequences as $A$ is increased above $1$. Sequential state
	points are numbered, the dotted diagonal line is the identity
	line $X_{n+1} = X_n$, the solid line is the map of the
	uncontrolled system with slope $A = 2.1$, the fixed point
	$\xi^*$ is denoted by the solid triangle, and the dot-dash
	lines show the system map when perturbed by control
	interventions. (a) $\beta=1.5$ results in an unstable $01^\infty$
	control sequence. (b) $\beta=2.5$; unstable $01^1$ sequence. (c)
	$\beta=3.5$; converging $01^1$ sequence. (d) $\beta=4.5$; unstable
	$010^\infty$ sequence. Note that axes from different panels are not
	necessarily scaled the same.}

\label{fig:x_return_a_above_1}
\end{figure}

\begin{figure}[!ht]

	\caption{Stability zones of restricted control
	(Eq.~\ref{eqn:restlin}) for $A>1$. The zone of semi-stability
	(shaded region; denoted $01^1$) and the different zones of
	instability are separated by the curves $\beta=A$,
	$\beta=1+A$, and $\beta=A^2/(A-1)$. The annotated open circles
	\textbf{a}--\textbf{d} correspond to the control parameters
	for panels (a)--(d) of Fig.~\ref{fig:x_return_a_above_1}.}

\label{fig:beta_vs_a_above_1}
\end{figure}

\begin{figure}[!ht]

	
	\caption{Control sequences observed in the rabbit heart
	experiments of Ref.~[21].  The first control perturbation in
	each panel is indicated by an arrow. (a) An unstable $01^1$
	sequence for $\alpha = 3.3$. (b) The same preparation with
	$\alpha = 5.0$. In this case the control begins in a stable
	$001^1$ sequence and shifts to a stable $01^1$ sequence.  (c)
	A stable $001^2$ sequence, shifting to a stable $01^2$
	sequence, and returning to a stable $001^2$ sequence in a
	different preparation with $\alpha = 2.5$.}

\label{fig:hall_97a_fig}
\end{figure}

\begin{figure}[!ht]

	\caption{Adaptive control of the quadratic map of
	Eq.~\ref{eqn:quadmap}. Control was activated during the
	intervals labelled with a ``C''. Iterates 125-375, 625-875,
	and 1125-1375 show control of period-2 ($\lambda_0=3.30$),
	period-4 ($\lambda_0=3.52$), and chaotic dynamics
	($\lambda_0=3.65$), respectively.}

\label{fig:quadcon}
\end{figure}

\begin{figure}[!ht]

	\caption{Adaptive control of a drifting quadratic map
	(Eq.~\ref{eqn:quadmap}). The baseline value of
	$\lambda_0=3.00$ was incremented by 0.001 each iterate
	($\lambda_{n+1}=\lambda_n+0.001$), causing a slow drift in the
	system. Control was activated from $250\leq n \leq 750$
	(labelled with a ``C''). During this period, $X^*$ drifted
	from $X^*=0.692$ to $X^*=0.736$. The adaptive algorithm
	tracked the drifting fixed point as the system moved into the
	chaotic regime.}

\label{fig:quadrift}
\end{figure}

\begin{figure}[!ht]

	\caption{Control of the cubic map of
	Eq.~\ref{eqn:cubemap} with with $\lambda_0 = 1.0$. Control was
	activated during the intervals labelled with a ``C''. Iterates
	125-375 and 625-875 show control of period-2 ($m=2.2$) and
	chaotic dynamics ($m=2.7$), respectively.}

\label{fig:cubecon}
\end{figure}

\begin{figure}[!ht]

	\caption{Adaptive control of a drifting cubic map
	(Eq.~\ref{eqn:cubemap}). The baseline value of $m_0=2.0$ was
	incremented by 0.001 each iterate ($m_{n+1}=m_n+0.001$),
	causing a slow drift in the system. Control was activated from
	$250\leq n \leq 750$ (labelled with a ``C'').  The fixed point
	location does not change for the drifting cubic map, but the
	system drifted into the chaotic regime by the end of the
	control period.}

\label{fig:cubedrift}
\end{figure}

\clearpage

\newcounter{fig_counter}
\setcounter{fig_counter}{0}

\newpage
\enlargethispage*{1000pt}
\noindent

	\begin{minipage}[b]{2.25in}
		\centerline{\includegraphics[width=2.25in,
		  bb=35 130 525 660,clip]{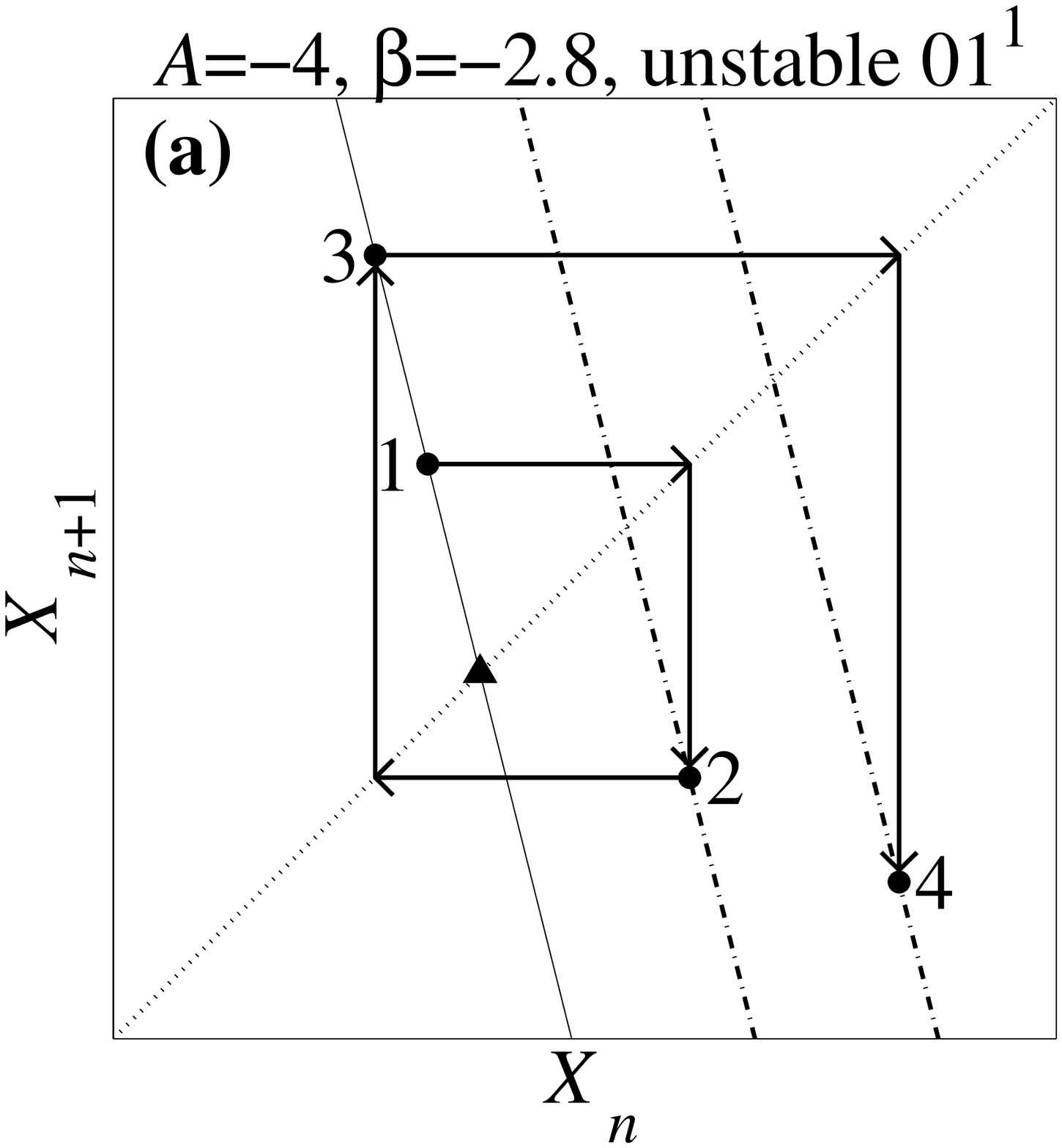}}
	\end{minipage}
	\begin{minipage}[b]{2.25in}
		\centerline{\includegraphics[width=2.25in,
		bb=35 130 525 660,clip]{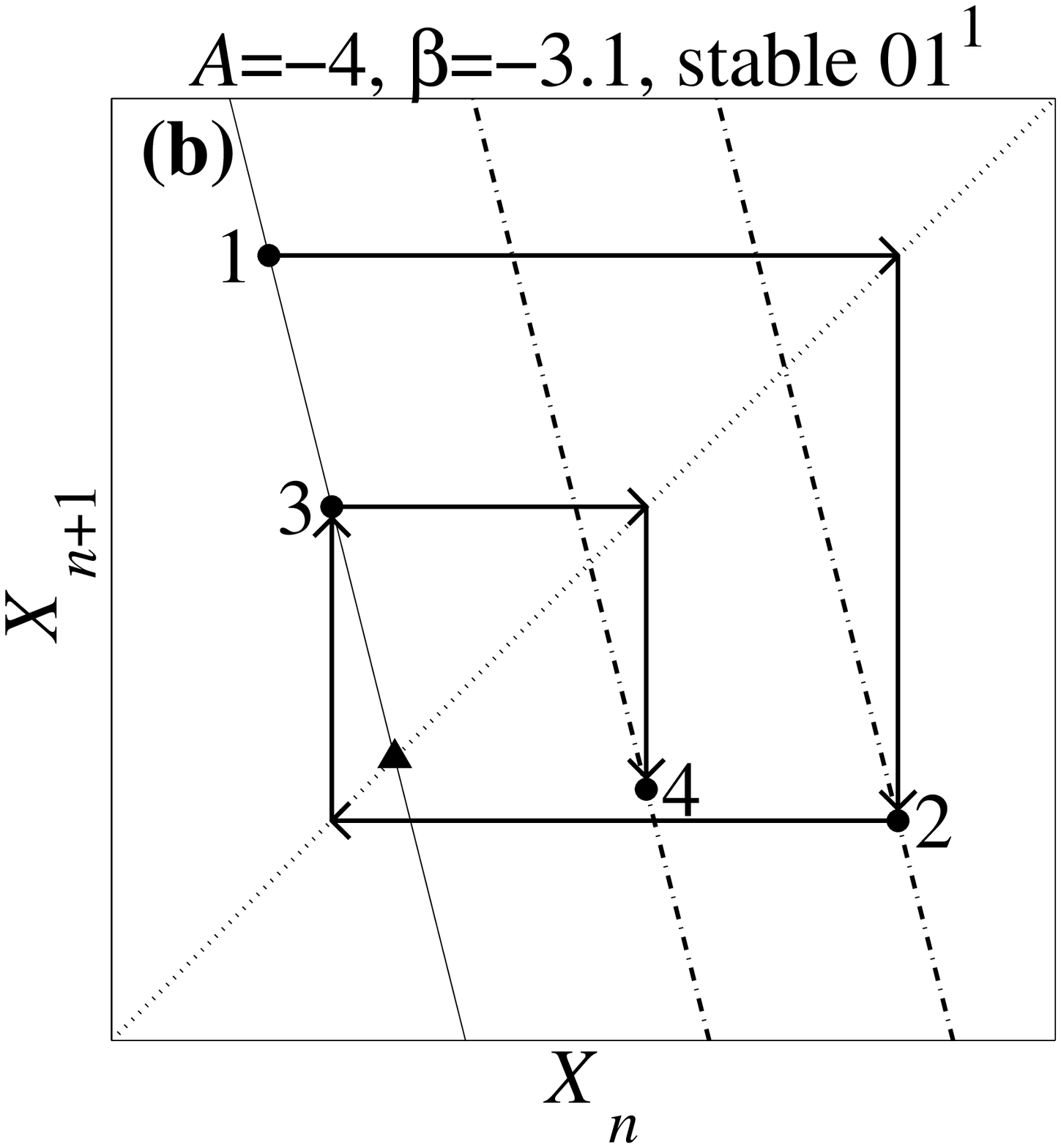}}
	\end{minipage}

	\begin{minipage}[b]{2.25in}
		\centerline{\includegraphics[width=2.25in,
		bb=35 130 525 660,clip]{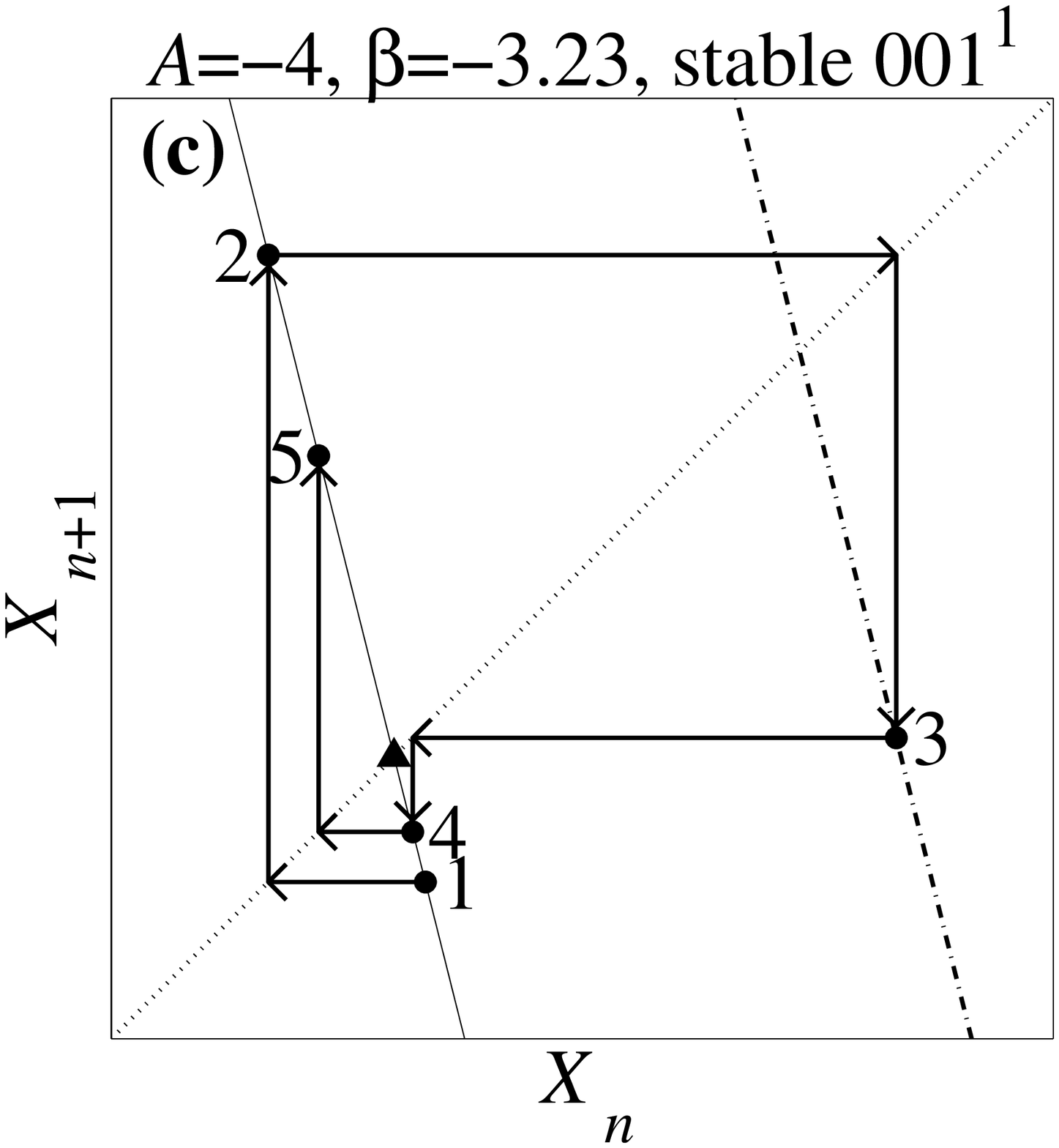}}
	\end{minipage}
	\begin{minipage}[b]{2.25in}
		\centerline{\includegraphics[width=2.25in,
		bb=35 130 525 660,clip]{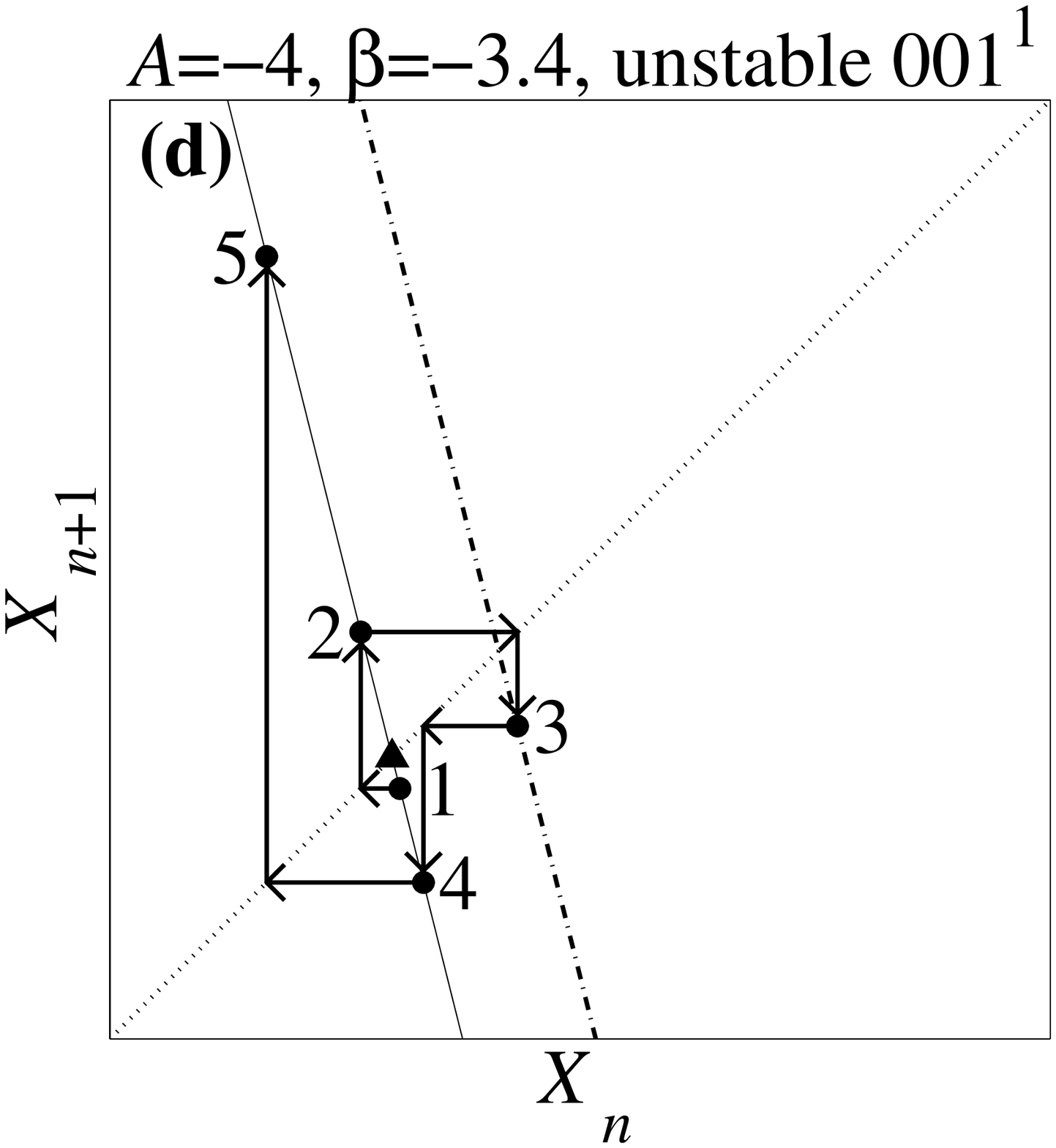}}
	\end{minipage}

	\begin{minipage}[b]{2.25in}
		\centerline{\includegraphics[width=2.25in,
		bb=35 130 525 660,clip]{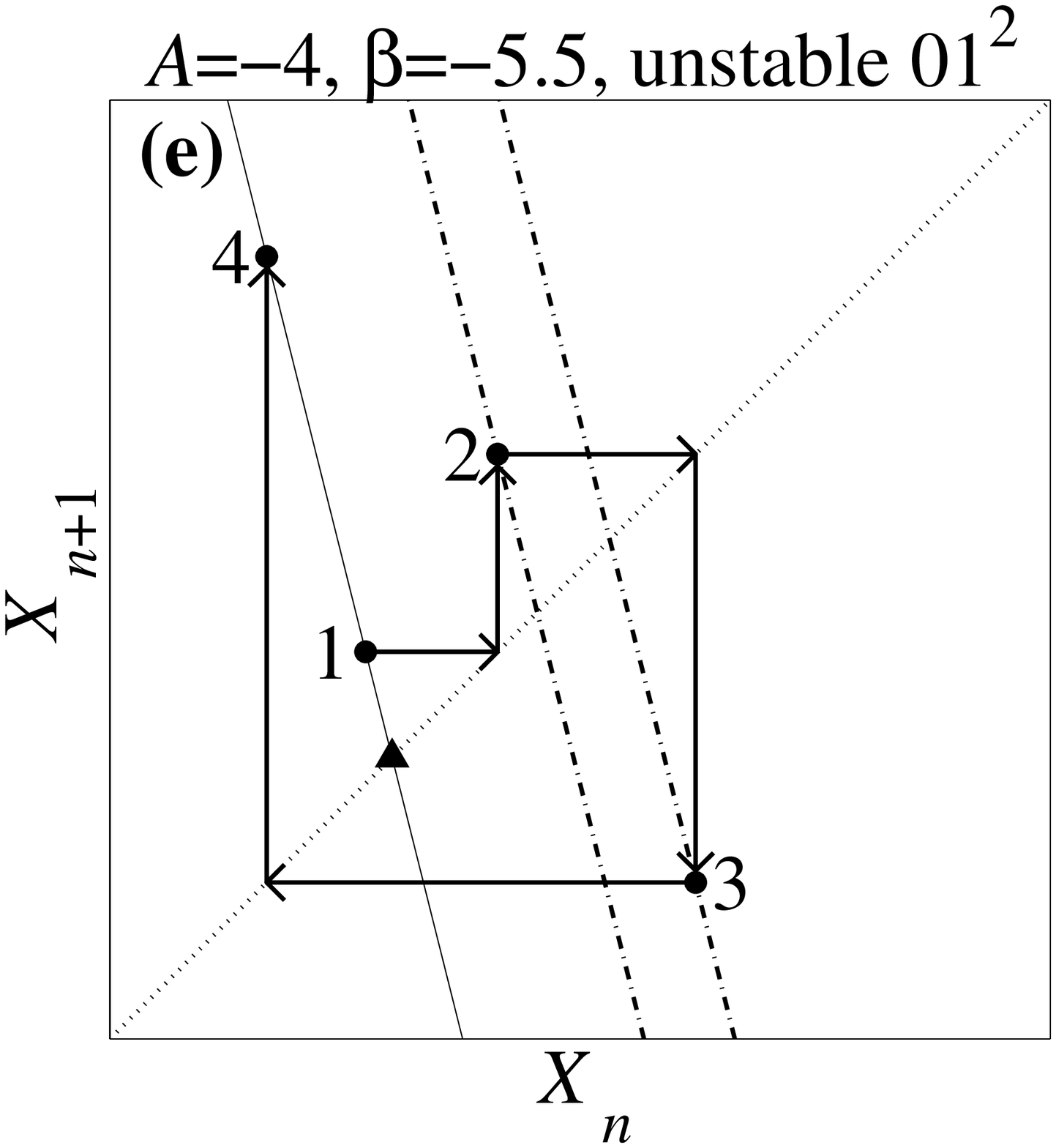}}
	\end{minipage}
	\begin{minipage}[b]{2.25in}
		\centerline{\includegraphics[width=2.25in,
		bb=35 130 525 660,clip]{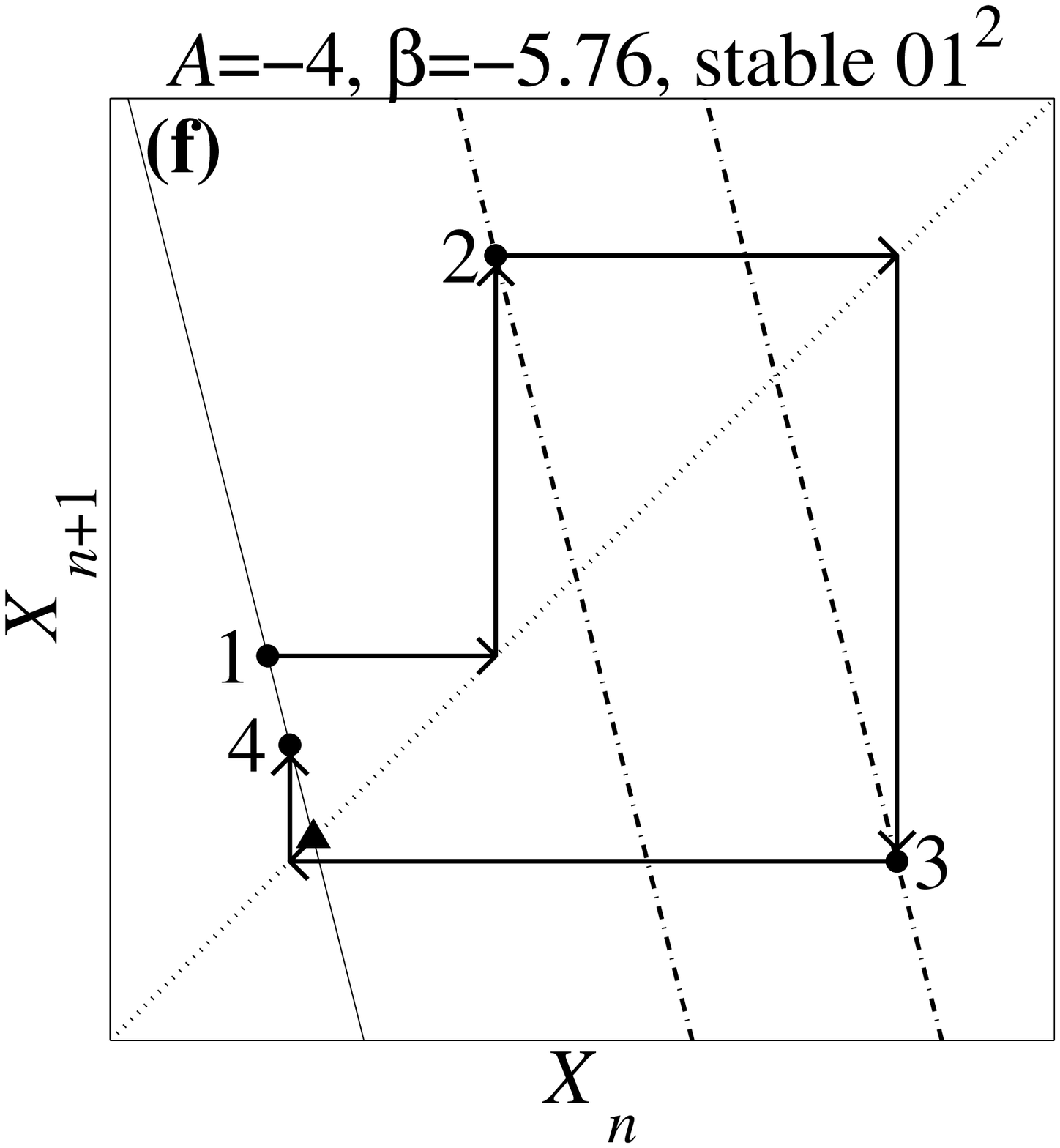}}
	\end{minipage}

	\begin{minipage}[b]{2.25in}
		\centerline{\includegraphics[width=2.25in,
		bb=35 130 525 660,clip]{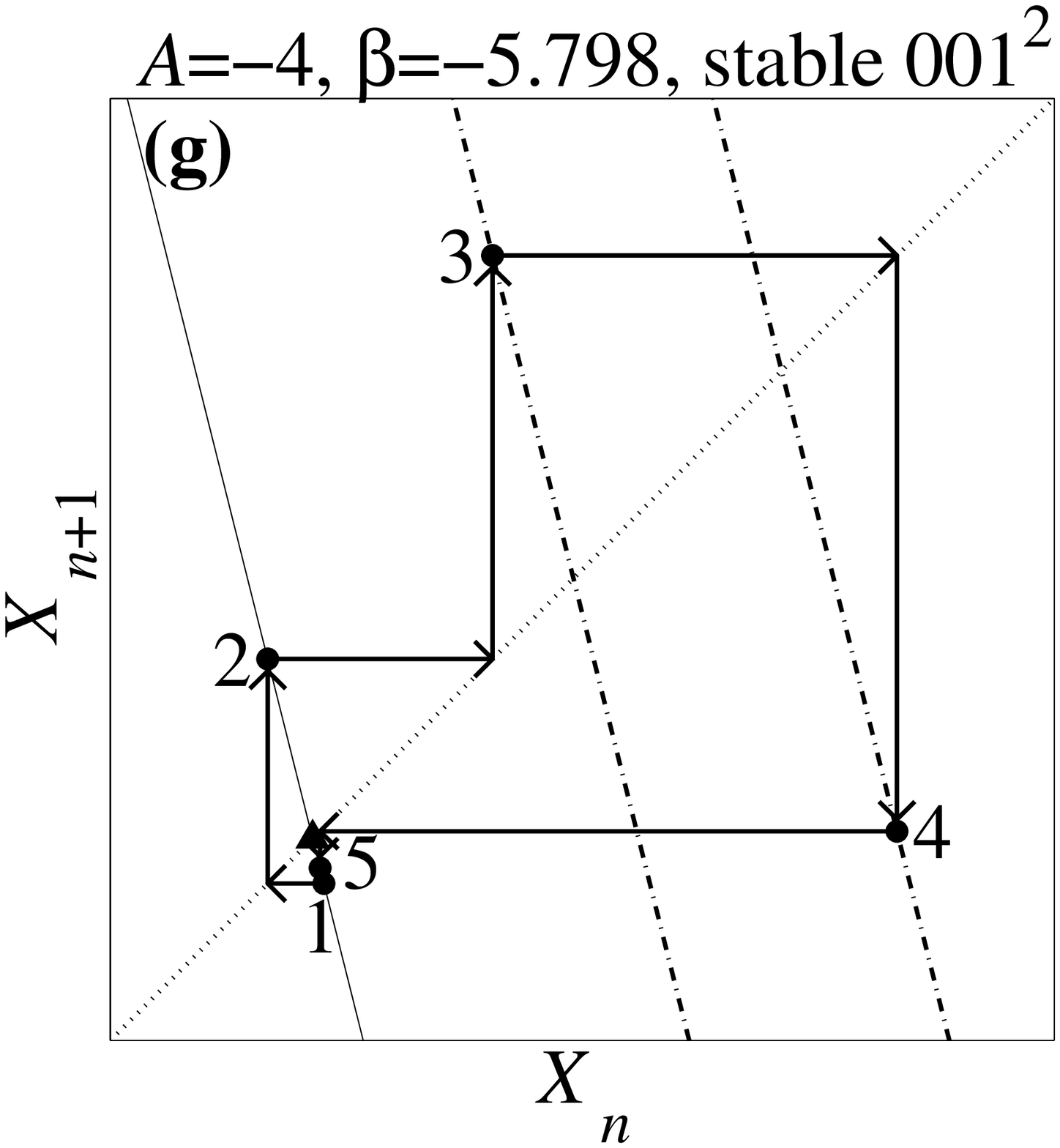}}
	\end{minipage}
	\begin{minipage}[b]{2.25in}
		\centerline{\includegraphics[width=2.25in,
		bb=35 130 525 660,clip]{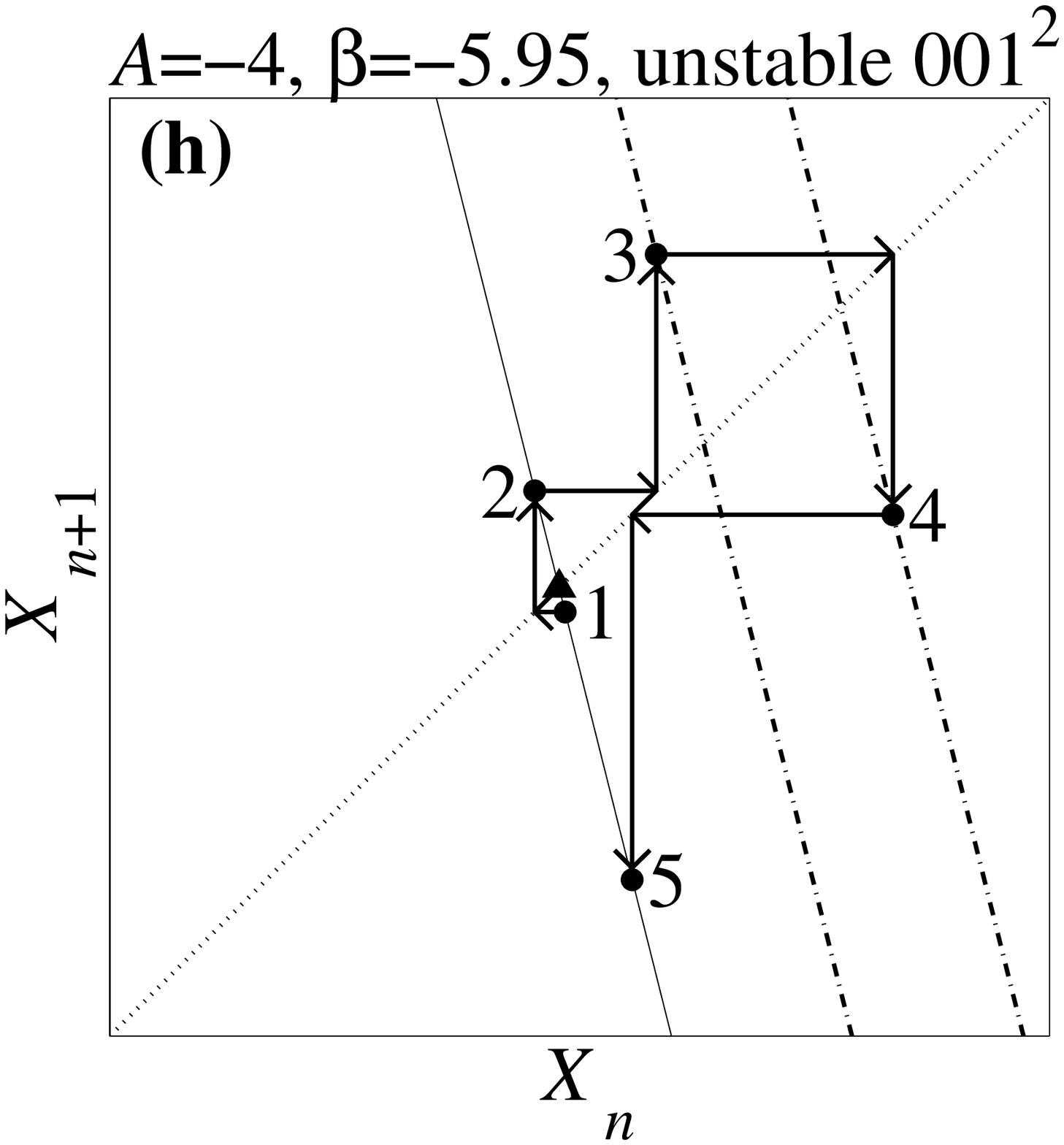}}
	\end{minipage}

\stepcounter{fig_counter}

\vspace*{-5.5in}
\begin{center}
\hspace*{5.25in}\Large Fig. \arabic{fig_counter}\\
\end{center}

\newpage
\enlargethispage*{1000pt}
\centerline{\includegraphics[height=7.5in]{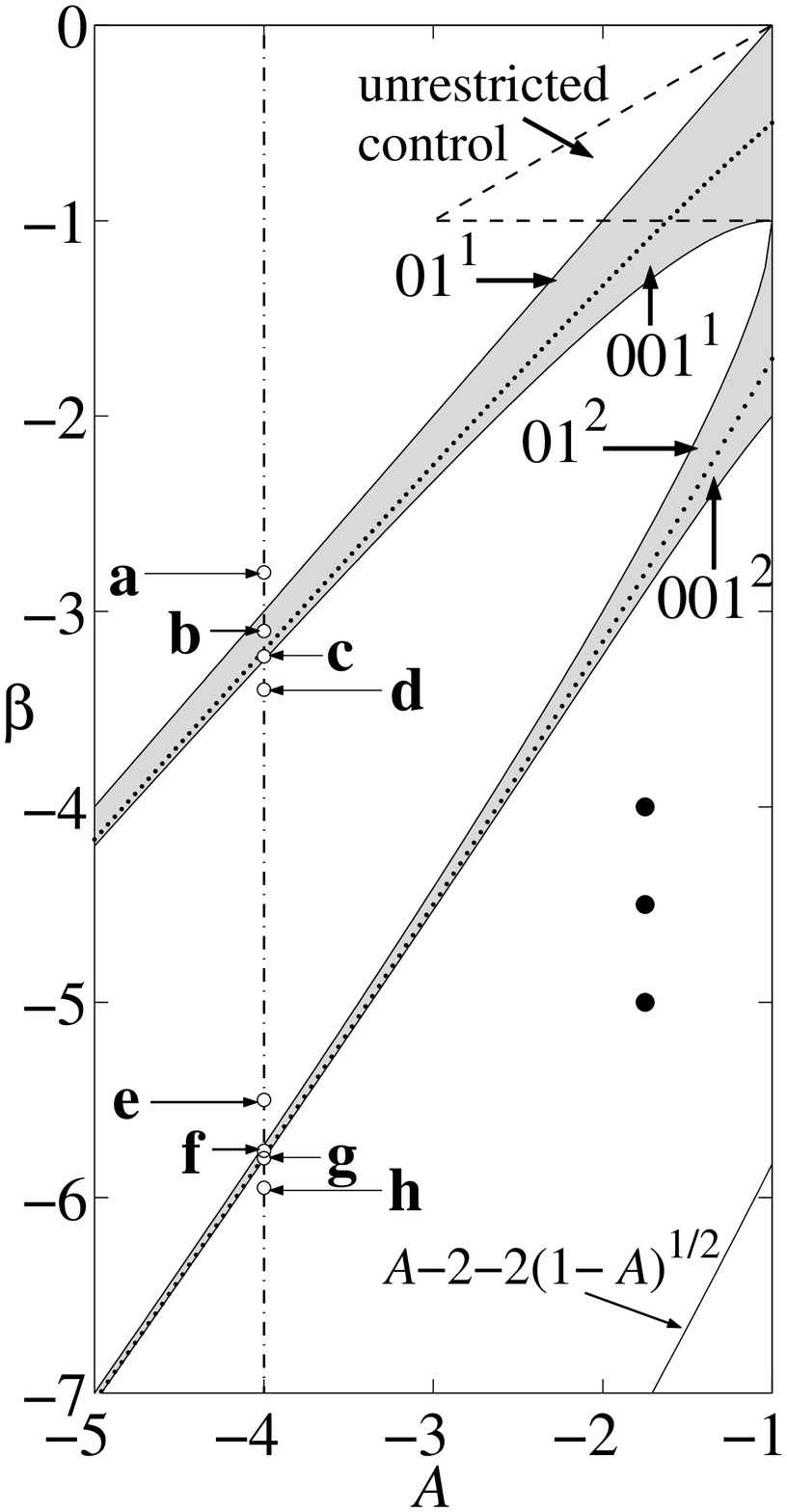}}
\stepcounter{fig_counter}
\begin{center}
\Large Fig. \arabic{fig_counter}\\
\end{center}

\newpage
\enlargethispage*{1000pt}
\noindent

	\begin{minipage}[b]{2.25in}
		\centerline{\includegraphics[width=2.25in,
		  bb=35 130 525 660,clip]{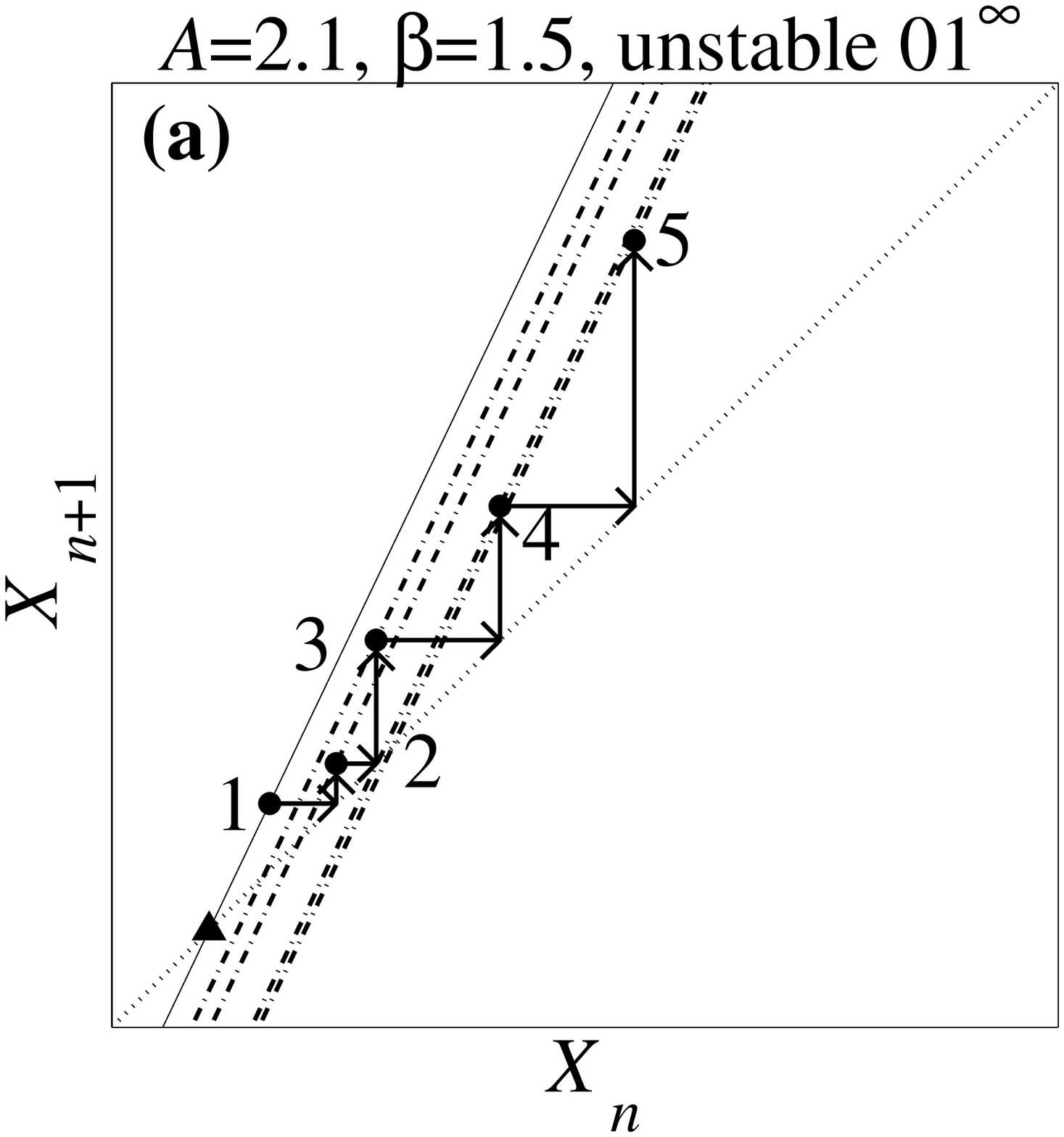}}
	\end{minipage}
	\begin{minipage}[b]{2.25in}
		\centerline{\includegraphics[width=2.25in,
		bb=35 130 525 660,clip]{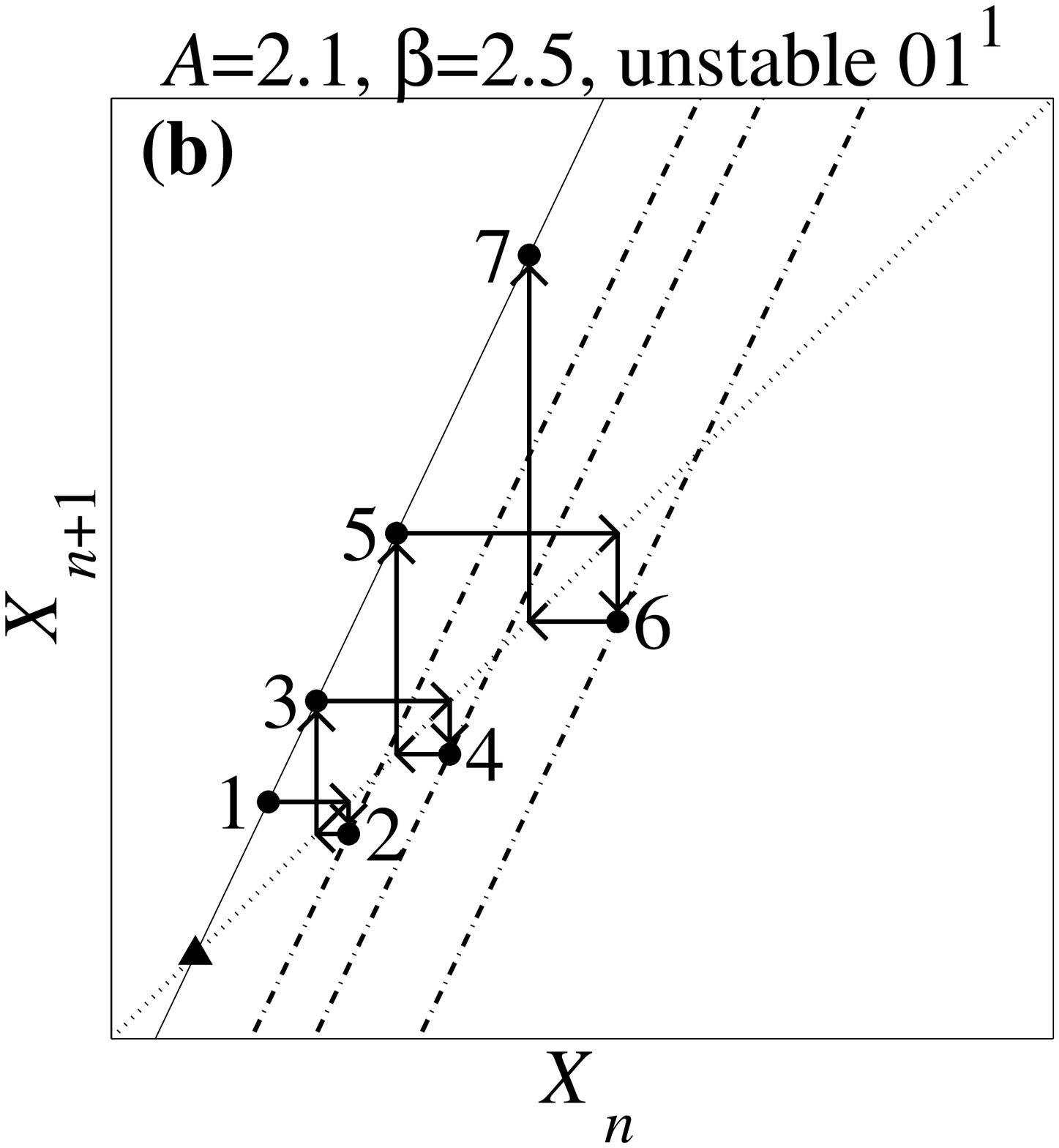}}
	\end{minipage}

	\begin{minipage}[b]{2.25in}
		\centerline{\includegraphics[width=2.25in,
		bb=35 130 525 660,clip]{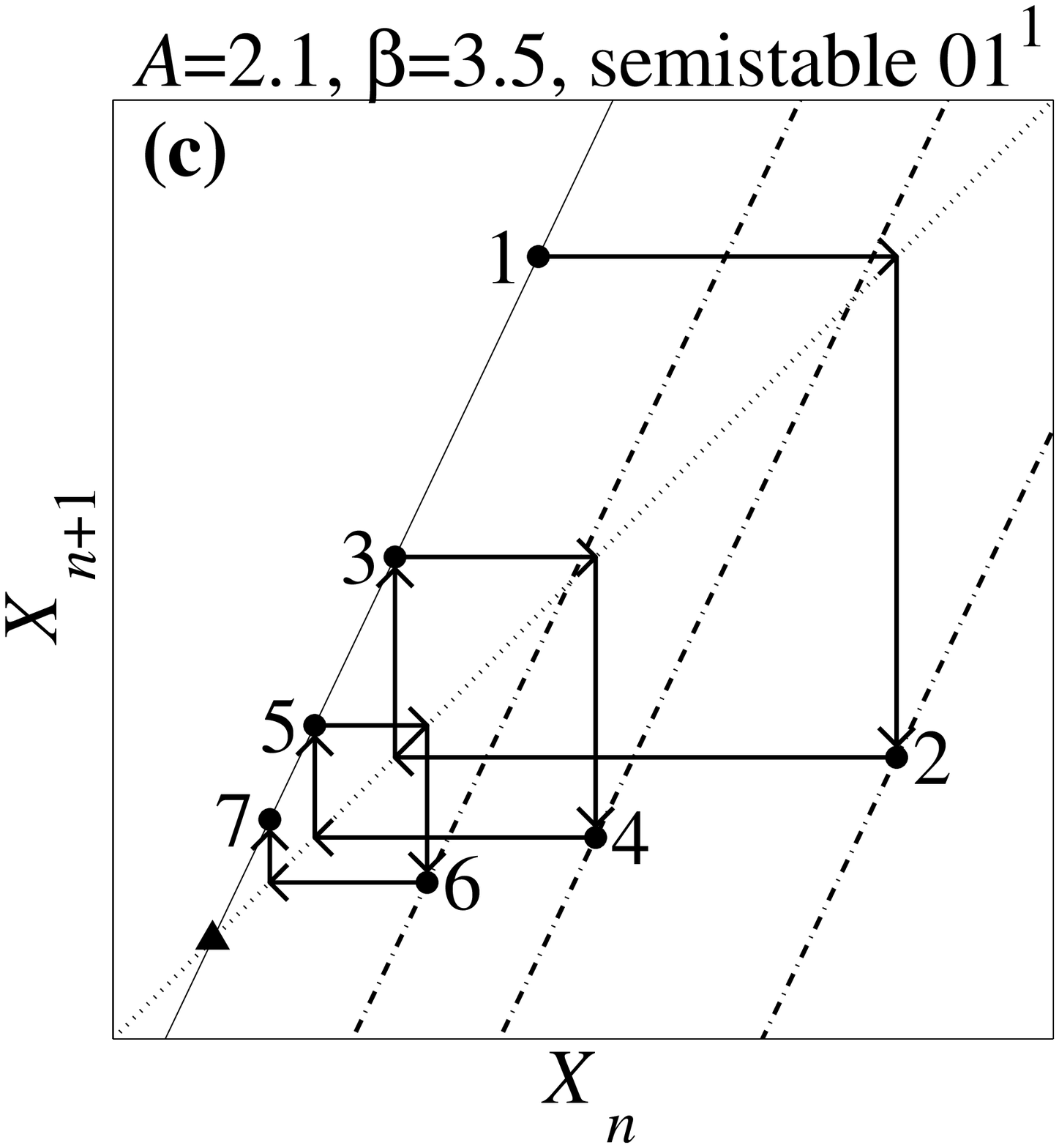}}
	\end{minipage}
	\begin{minipage}[b]{2.25in}
		\centerline{\includegraphics[width=2.25in,
		bb=35 130 525 660,clip]{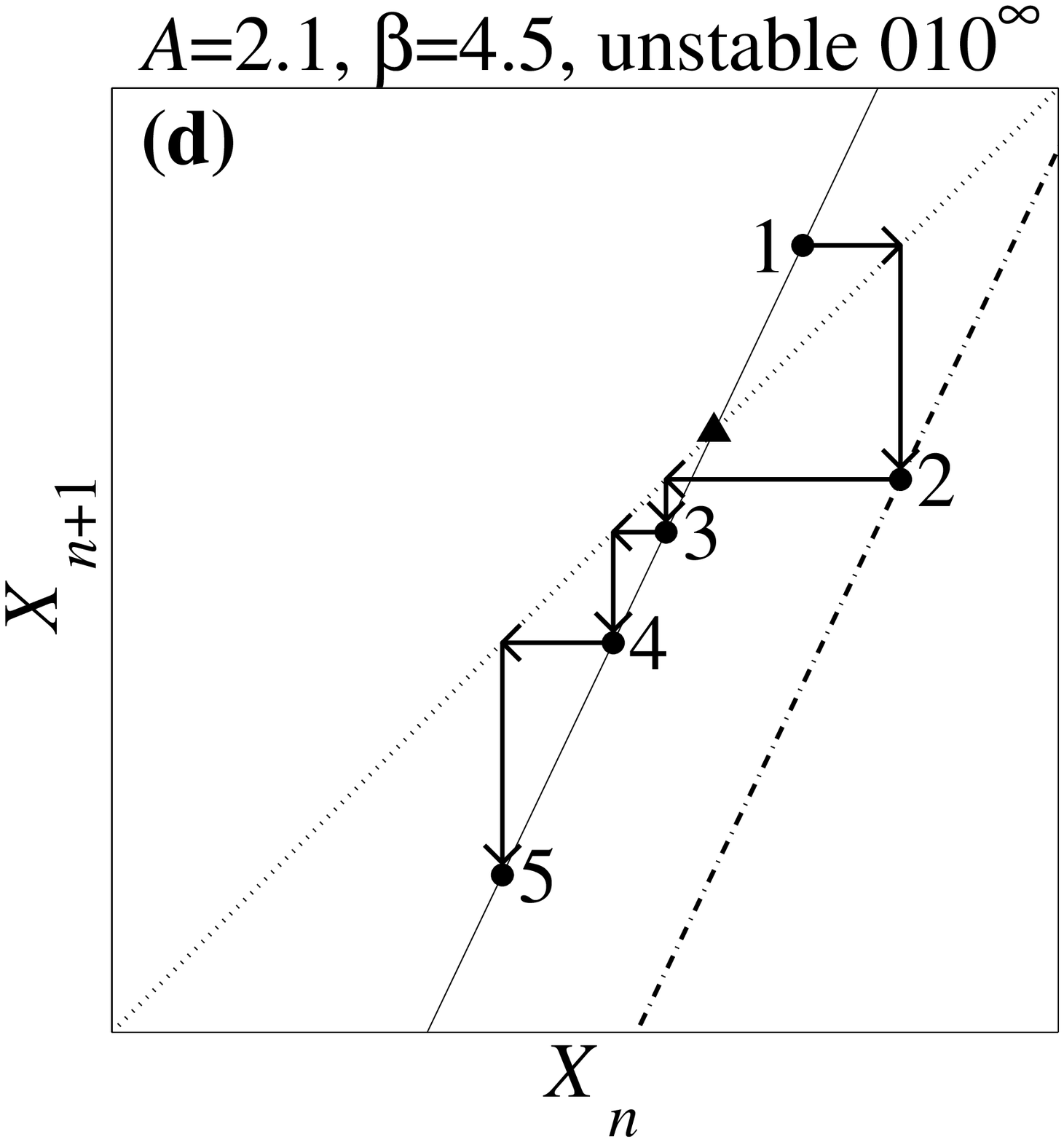}}
	\end{minipage}

\stepcounter{fig_counter}

\begin{center}
\hspace*{-1.25in}\Large Fig. \arabic{fig_counter}\\
\end{center}

\newpage
\enlargethispage*{1000pt}
\centerline{\includegraphics[height=7.5in]{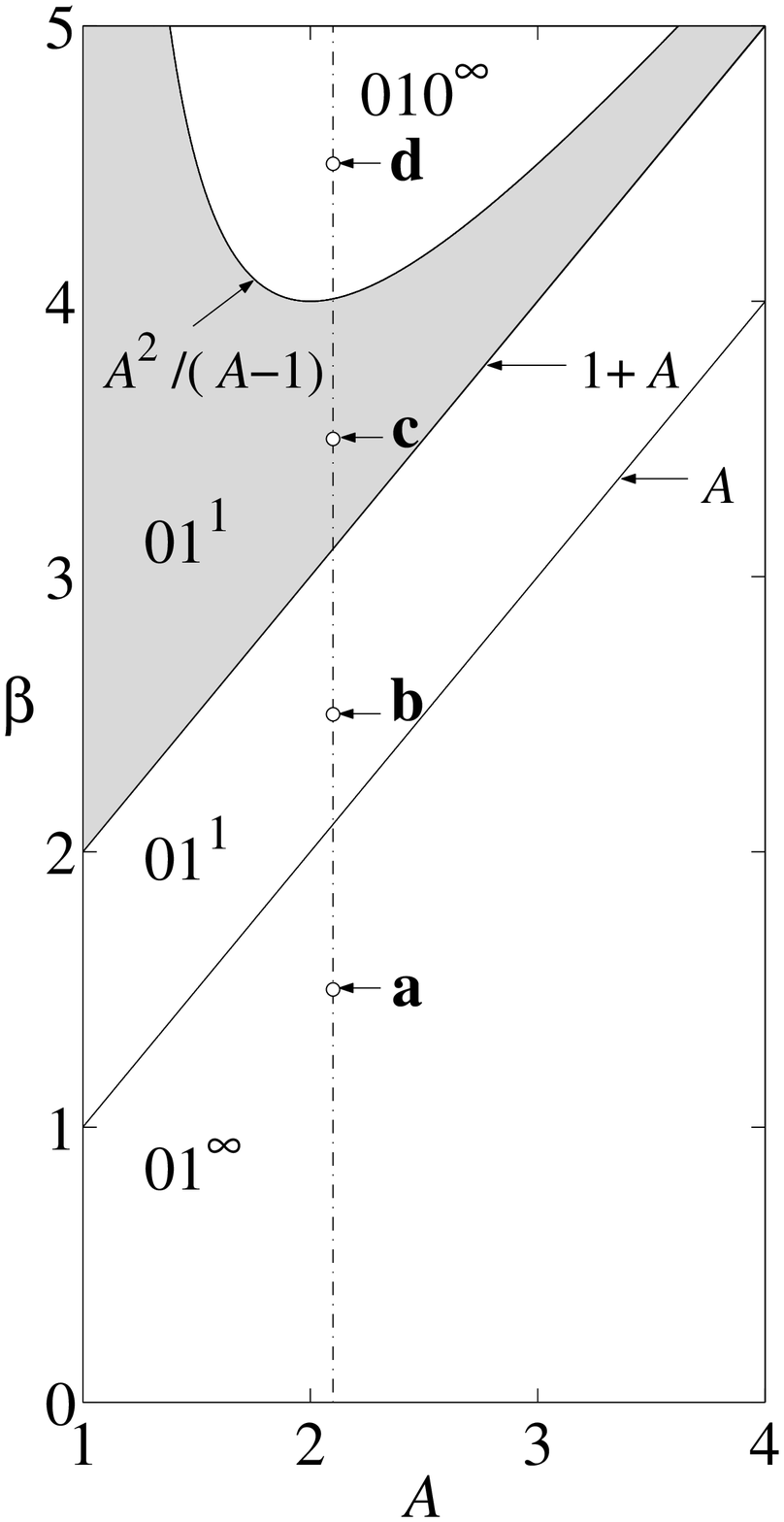}}
\stepcounter{fig_counter}
\begin{center}
\Large Fig. \arabic{fig_counter}\\
\end{center}

\newpage
\enlargethispage*{1000pt}
\centerline{\includegraphics[height=8.5in]{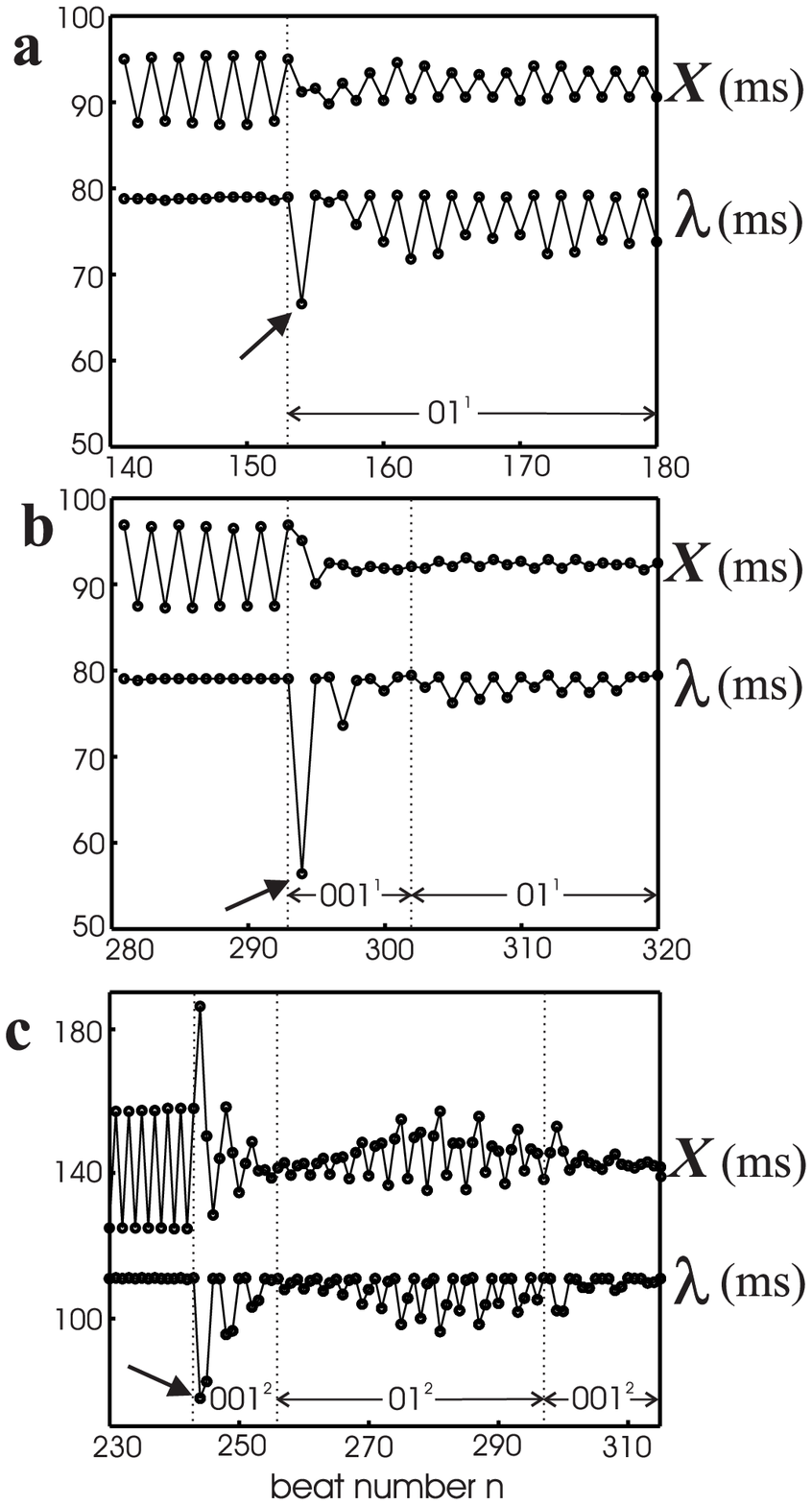}}
\stepcounter{fig_counter}
\begin{center}
\Large Fig. \arabic{fig_counter}\\
\end{center}

\newpage
\enlargethispage*{1000pt}
\vspace*{0.75in}
\centerline{\includegraphics[width=5.5in,clip]{./hall_00a_fig6.eps}}
\stepcounter{fig_counter}
\begin{center}
\Large Fig. \arabic{fig_counter}\\
\end{center}

\newpage
\enlargethispage*{1000pt}
\vspace*{0.75in}
\centerline{\includegraphics[width=5.5in,clip]{./hall_00a_fig7.eps}}
\stepcounter{fig_counter}
\begin{center}
\Large Fig. \arabic{fig_counter}\\
\end{center}

\newpage
\enlargethispage*{1000pt}
\vspace*{0.75in}
\centerline{\includegraphics[width=5.5in,clip]{./hall_00a_fig8.eps}}
\stepcounter{fig_counter}
\begin{center}
\Large Fig. \arabic{fig_counter}\\
\end{center}

\newpage
\enlargethispage*{1000pt}
\vspace*{0.75in}
\centerline{\includegraphics[width=5.5in,clip]{./hall_00a_fig9.eps}}
\stepcounter{fig_counter}
\begin{center}
\Large Fig. \arabic{fig_counter}\\
\end{center}

\end{document}